\documentclass[conference]{IEEEtran}
\IEEEoverridecommandlockouts
\usepackage{cite}
\usepackage{amsmath,amssymb,amsfonts}
\usepackage{algorithmic}
\usepackage{graphicx}
\usepackage{textcomp}
\usepackage{xcolor}
\usepackage[colorinlistoftodos]{todonotes}
\def\BibTeX{{\rm B\kern-.05em{\sc i\kern-.025em b}\kern-.08em
    T\kern-.1667em\lower.7ex\hbox{E}\kern-.125emX}}

\newcommand\copyrighttext{%
  \footnotesize \textcopyright 2022 IEEE. Personal use of this material is permitted.
  Permission from IEEE must be obtained for all other uses, in any current or future 
  media, including reprinting/republishing this material for advertising or promotional 
  purposes, creating new collective works, for resale or redistribution to servers or 
  lists, or reuse of any copyrighted component of this work in other works. 
  \linebreak Preprint submitted to the 27th IEEE International Conference on Emerging Technologies and Factory Automation, ETFA 2022.
  }
\newcommand\copyrightnotice{%
\begin{tikzpicture}[remember picture,overlay]
\node[anchor=south,yshift=10pt] at (current page.south) {\fbox{\parbox{\dimexpr\textwidth-\fboxsep-\fboxrule\relax}{\copyrighttext}}};
\end{tikzpicture}%
}

\begin{document}

\title{A Civil Protection Early Warning System to Improve the Resilience of Adriatic-Ionian Territories to Natural and Man-made Risk
\thanks{This work has been supported by the project "Establishment of 'TRANSnational Civil Protection EARLY WARNING System' to improve the resilience of Adrion territories to natural and man-made risks" - TransCPEarlyWarning 979,  implemented under the INTERREG ADRION programme, co-financed by the European Regional Development Fund (ERDF), Instrument for Pre-Accession Assistance (IPA), and national sources.}
}

\author{
\IEEEauthorblockN{Agorakis Bompotas}
\IEEEauthorblockA{\textit{Industrial Systems Institute} \\
\textit{Athena Research Center}\\
Patras, Greece \\
ORCID: 0000-0002-6063-8562}
\and
\IEEEauthorblockN{Christos Anagnostopoulos}
\IEEEauthorblockA{\textit{Industrial Systems Institute} \\
\textit{Athena Research Center}\\
Patras, Greece \\
ORCID: 0000-0002-7998-5708}
\and
\IEEEauthorblockN{Athanasios Kalogeras}
\IEEEauthorblockA{\textit{Industrial Systems Institute} \\
\textit{Athena Research Center}\\
Patras, Greece \\
ORCID: 0000-0001-5914-7523}
\and
\IEEEauthorblockN{Georgios Kalogeras}
\IEEEauthorblockA{\textit{Industrial Systems Institute} \\
\textit{Athena Research Center}\\
Patras, Greece \\
ORCID: 0000-0001-9831-4291}
\and
\IEEEauthorblockN{Georgios Mylonas}
\IEEEauthorblockA{\textit{Industrial Systems Institute} \\
\textit{Athena Research Center}\\
Patras, Greece \\
ORCID: 0000-0003-2128-720X}
\and
\IEEEauthorblockN{Kyriakos Stefanidis}
\IEEEauthorblockA{\textit{Industrial Systems Institute} \\
\textit{Athena Research Center}\\
Patras, Greece \\
ORCID: 0000-0002-2090-2218}
\and
\IEEEauthorblockN{Christos Alexakos}
\IEEEauthorblockA{\textit{Industrial Systems Institute} \\
\textit{Athena Research Center}\\
Patras, Greece \\
ORCID: 0000-0002-8932-6781}
\and
\IEEEauthorblockN{Miranda Dandoulaki}
\IEEEauthorblockA{\textit{Industrial Systems Institute} \\
\textit{Athena Research Center}\\
Patras, Greece \\
email: mdand@isi.gr}
}

\maketitle

\copyrightnotice

\begin{abstract}
We are currently witnessing an increased occurrence of extreme weather events, causing a great deal of disruption and distress across the globe. In this setting, the importance and utility of Early Warning Systems is becoming increasingly obvious. In this work, we present the design of an early warning system called TransCPEarlyWarning, aimed at seven countries in the Adriatic-Ionian area in Europe. The overall objective is to increase the level of cooperation among national civil protection institutions in these countries, addressing natural and man-made risks from the early warning stage and improving the intervention capabilities of civil protection mechanisms. The system utilizes an innovative approach with a lever effect, while also aiming to support the whole system of Civil Protection. 
\end{abstract}


\section{Introduction}

Climate change is marked by its impact in indicators such as the average temperature, precipitation, ocean acidification, sea-level rise and extreme weather conditions, and is manifested by events such as droughts, floods, hurricanes, severe storms, heatwaves, wildfires, cold spells and landslides \cite{united2019climate}. The trend is that the frequency of several types of extreme weather events will increase during the 21st century, including heat waves, heavy precipitation, droughts, and tropical cyclones, resulting in an increasing frequency and severity of relevant hazards \cite{banholzer2014impact}.

The Adriatic-Ionian area in Europe is prone to different types of natural and man-made hazards of increased severity and frequency, following the global norm. National Civil Protection (CP) Systems deal with the prediction, prevention and management of emergencies. In this context, increasing the level of cooperation between the National Civil Protection Systems in the different countries in the area, and exchanging good practices related to the field, can help improve their efficiency and overall response to hazards.

The TransCPEarlyWarning project \cite{kalogeras2020establishment} approaches the aforementioned challenges by equipping Adriatic Ionian Civil Protection with a transnational Early Warning System that intends to increase overall efficiency, interoperability and homogeneity in handling natural risks at their Early Warning stage.  Its core element is the TransCPEarlyWarning Civil Protection Early Warning Platform, which serves the purpose of offering a web-enabled semantically enriched focal point of reference for the Civil Protection stakeholders in Adriatic-Ionian (ADRION) territories, enabling Civil Protection Early Warning process monitoring, integration of existing Early Warning systems and experimentation with  Artificial Intelligence/Machine Learning (AI/ML) algorithms.

The project follows a three pillar approach in order to deliver its vision. At the governance level, it studies the existing Civil Protection Early Warning regulatory systems and legislation in the different countries of the Adriatic Ionian area, aiming at a common model and specific recommendations for change. At the innovation level it introduces the aspect of a common Early Warning Platform (EWP) for the area. Finally, at the EU integration level it addresses the need for a unifying approach for the area that comprises both EU member and non-member states. This paper will focus on the design, and some implementation aspects, of the Early Warning Platform as well as the integration with external civil protection services.

The rest of the paper is structured as follows. Section II provides background information and previous work. Section III elaborates on the TransCPEarlyWarning Platform design methodology. Section IV presents the platform conceptual architecture. Section V deals with integration of external systems. Section VI refers to the platform security specifications. Section VII discusses the TransCPEarlyWarning platform implementation. Finally, section VIII provides conclusions.

\section{Previous Work}

Civil Protection in Europe first emerged in the 1980s, out of concerns over catastrophic events and their impact, kick-starting cooperation between France and Italy towards assessing hazards and risks that they presented to the public \cite{gaetani2009structure}. A transnational approach for Civil Protection is often a necessity, since different countries can face the same hazards across borders; there is also in many cases a need for emergency assistance between countries, and particular expertise in the field can be transferred between different them to improve the handling of catastrophic events and mitigate their impact. 

Such needs can be hindered in practice by a number of elements, including  differences in terminology used among countries, multiple legal frameworks, regulations and plans, main stakeholders and technological solutions. Depending on the hazard addressed, there are several ongoing international efforts and systems that attempt to homogenize processes and systems.

The main components of a Civil Protection Early Warning System are related to (i) disaster risk knowledge, (ii) detection, monitoring, analysis and forecasting of hazards and possible consequences, (iii) dissemination and communication, and (iv) preparedness to respond to warnings \cite{unisdr2021terminology}. Some important aspects of Civil Protection Early Warning Systems are the hazards they address, their main goal, their spatial scale, the levels of administration and involved  stakeholders, and the technologies and solutions that are used.

\subsection{Existing International Early Warning Systems}

EM-DAT \cite{em2019dat} is a free-to-use, global, and fully searchable database containing information about natural and technological disasters from 1900 to present. It contains core data on the occurrence and effect of more than 21,000 distinct cases. The main purpose of EM-DAT is to assist humanitarian action at both national and international level, rationalize decision-making for disaster preparedness, and provide an objective basis for vulnerability assessment and priority setting. EM-DAT provides two different tools, namely the Querying tool and the Mapping tool: the Querying tool provides filters for refining the search of the database, relevant to disaster classification, location and time, while the Mapping tool provides a GUI for querying disasters and returns a geographical summary of the results.

Copernicus \cite{copernicus} is an EU programme aimed at developing European information services based on satellite Earth Observation and in situ (non space) data. The objective of Copernicus is to monitor and forecast the state of the environment on land, sea and in the atmosphere, in order to support climate change mitigation and adaptation strategies, the efficient management of emergency situations and the improvement of the security of every citizen. Copernicus is a user-driven programme and the information services provided are available to its users, mostly public authorities, on a full, open and free-of-charge basis. Copernicus Early Warning and Monitoring offers critical geospatial information at European and global level through continuous observations and forecasts for floods, droughts and forest fires.

The European Forest Fire Information System (EFFIS) \cite{san2012comprehensive} supports the services in charge of the protection of forests against fires in the EU countries and provides the European Commission services and the European Parliament with updated and reliable information on  wildfires in Europe since 2000. It is related to the development and implementation of advanced methods for the evaluation of forest fire danger and mapping of burnt areas at the European scale. The most important tool provided by EFFIS is the current situation viewer, which provides up-to-date information on the fire season in the area of interest. The available information includes meteorological fire danger maps and specific weather forecast concerning fire related weather anomalies. Moreover, the system provides daily maps of forest fires and burnt area perimeters. Info is presented on an interactive map and useful graphs with relevant data are also available to the user. The Fire News application presents all fire related news from every country on an interactive map. The user can actually browse news for a specific country and area. The system also offers long-term monthly and seasonal fire weather forecasts, as well as data on demand by the end user.

The aim of the European Flood Awareness System (EFAS) \cite{thielen2009european} is to support preparatory measures before major flood events occur, particularly in large transnational river basins, and throughout Europe in general. It provides complementary, added-value information (e.g., probabilistic, medium range flood forecasts, flash flood indicators or impact forecasts) to the relevant national and regional authorities, while keeping Emergency Response Coordination Center informed about ongoing and possibly upcoming flood events across Europe. The key elements in the EFAS production chain are the meteorological forcing and land surface data, the hydrological models, and the EFAS forecasts and products. The meteorological forcing data comprise Historical hydro-meteorological time series records, Real-time hydro-meteorological observations, and Meteorological forecasts, while land surface data comprise all information necessary to set-up and implement the hydrological models. EFAS hydrological models transform the meteorological forcing forecasts into hydrological forecasts by mimicking the hydrological land processes by a set of equations: hydrological model LISFLOOD \cite{van2010lisflood} is primarily used for medium- and seasonal-range forecasts, whilst conceptual hydrological algorithms are used for the flash flood indicators. Finally, EFAS forecasts are produced by forcing the LISFLOOD hydrological model with a range of meteorological forecasts.

The Global Disaster Alert and Coordination System (GDACS) \cite{de2006global} is a cooperation framework between the United Nations and the European Commission. It includes disaster managers and disaster information systems worldwide and aims at filling the information and coordination gaps in the first phase after major disasters. Its main purpose is the provision of up-to-date, reliable, accurate and accessible information to the international early responder community after sudden onset disasters and the improvement of cooperation of international responders in the immediate aftermath of major natural, technological and environmental disasters. The primary service of GDACS is an integrated web-based system that combines critical disaster information systems under one umbrella. Natural disasters, such as earthquakes, tsunamis, floods, cyclones and droughts, all fall in the GDACS scope. GDACS integrates different tools including GDACS Disaster Alerts dissemination information to some 25,000 subscribers immediately following sudden-onset disasters, Virtual OSOCC (On-Site Operations Coordination Centre) for real-time information exchange and cooperation among all actors in the early phase of the event, and GDACS Satellite Mapping and Coordination System (SMCS) that provides communication and coordination for stakeholders to monitor and inform of completed, current and future mapping activities during emergencies. GDACS-SMCS main parts include satellite mapping coordination, satellite mapping overview reports, and live maps.

\subsection{AI/ML-based Early Warning Systems}

The recent developments in Artificial Intelligence, Machine Learning and Deep Learning (DL) have enabled us to apply such techniques to a wide range of application domains. Since we have also witnessed a rapid increase in the ways in which we are able to monitor the physical world, consequently there is a wealth of data available to be utilized for various purposes related to the world around us. In this context, the research community has turned its attention to utilizing such developments in order to develop early warning systems. Some examples of use cases for AI-based early warning systems that have surfaced in previous years are the following:

\begin{itemize}
\item \textit{Wildfires and smoke detection}: based on visual inputs, such as data from regular stationary cameras, imagery from thermal cameras or satellites, these systems can provide estimations as to whether what is seen in these visual data can be categorized as fire or smoke. Some of these systems even provide further categorizations, e.g., fire at an early stage, or provide additional information such as the direction of smoke based on previous input.
\item \textit{Floods}: based on on-site sensing input such as water level sensors, or satellite imagery, such systems can produce early warnings regarding the possibility of floods taking place within the following period. Similar techniques can be applied to accelerate the process of flood mapping as well and accelerate disaster relief planning.
\item \textit{Landslides}: based on sensor and weather data input, as well as satellite imagery, such systems can predict potential landslide occurrence in areas of interest.
\item \textit{COVID-19 pandemic}: based on mobile network and GPS data, together with other data such as street cameras, such systems have been used to predict the number of COVID-19 cases in the near future, to aid in making public health-related decisions.
\end{itemize}

Along with the appearance of such approaches to solve the problem of automatic fire and smoke detection, a number of initiatives have started to appear trying to tackle the issue of data inputs to be used to test such systems. The first of these initiatives began to surface around 2005, while their number began to increase at a faster rate after 2015. Since then, we have seen initiatives such as the Corsican Fire Database\cite{corsican-fire-db, corsican-website}, which provides an image dataset in multiple spectra with data annotations regarding e.g., the percentage of fire pixels in the image, the percentage of fire pixels covered by smoke, the level of texture of the fire area. Similarly, an example of a wildfire smoke dataset is provided by the ``Wildfire Smoke Dataset''~\cite{wildfire-smoke-dataset}, which includes annotated images of forest areas with smoke. Other, more recent, categories of datasets provided images from satellites, or even from UAVs while inspecting forest areas. Examples of images included in the aforementioned datasets are provided below.

Regarding the availability of relevant techniques and algorithms for this application use case, we have witnessed a range of approaches in this field. A comprehensive survey of relevant techniques in this domain is included in \cite{wildfire-ew-review}, where a large number of algorithms and techniques, utilizing data from terrestrial, airborne, and satellite-based systems used to detect fire and smoke, are presented.  A variety of approaches are being used among these systems, including computer vision-based ones, as well as ML/DL-based~\cite{satellite-wildfires, cnn-wildfire}. However, although the former approaches have been around for more years, the latter seem to be evolving faster and becoming more popular. 

\section{Platform Design Methodology}

The main goal of the TransCPEarlyWarning platform is to provide CP stakeholders with a holistic approach to Civil Protection management. First of all, it offers users extended management and monitoring capabilities of the relevant processes they are involved in. The platform offers stakeholders a single focal point of reference, where they can have access to the different information sources and systems utilized in their everyday routine with reference to forest fires and floods, the two main risks that are addressed in the current state of the platform integration. The nature of work of CP officers, requiring continuous process management even when out of office, demands creation of a web-enabled semantically-enriched platform, supplemented by access to such information from mobile devices, allowing access to critical process information on the go, from a securely designed and implemented environment. Finally, to promote innovation in the field of CP and experimentation based on open datasets and open-source code, relevant algorithms on hazard detection are offered as services to Civil protection stakeholders. With the above general principles in mind, the methodology for Platform design is presented in the next chapter.

\subsection{Specifications Elicitation Methodology}

The methodology we followed for the collection of the platform requirements and specifications utilized several different best practices. Firstly, an analysis of existing regulatory frameworks and systems relevant to ADRION Civil Protection Early Warning gave an understanding of the as-is situation in the Adriatic Ionian macro-region. Then, an additional survey with experts from the Adrion area countries having CP responsibilities provided insight into existing Civil Protection Early Warning processes. Brainstorming among the relevant stakeholders resulted in the collection of information related to user expectations for the envisaged platform. The final step involved gathering feedback from Focus Groups related to the Civil Protection Early Warning everyday routine and to the potential of the envisaged platform. Two focus groups were formed:
  \begin{itemize}
      \item The first focus group of 20 stakeholders was associated with the organizations that have responsibilities in Civil Protection at regional/local level, and targeted specifically their day-to-day operations. We collected information on the different systems/tools/websites utilized by the stakeholders in their everyday operations. 
      \item The second focus group of 14 stakeholders includes the members of the innovative transnational Network for Civil Protection Early Warning established in the context of the project. This group focused on collection of requirements related to the day-to-day operations of Civil Protection stakeholders, as well as their expectations from the envisaged Civil Protection Early Warning platform. The second group is organized in subgroups in different countries involving sessions/interviews with stakeholders.
  \end{itemize}

Based on the above methodology, the platform specifications presented in the next sections were defined.

\section{Platform Conceptual Architecture}

The TransCPEarlyWarning Civil Protection Early Warning Platform provides a number of functionalities, including: (i) the design, execution and monitoring of Civil Protection Early Warning procedures, (ii) the provision of access to different tools/sources of information related to Civil Protection, for facilitating the everyday routine of Civil Protection stakeholders (as defined during surveys with them), (iii) the experimentation with open datasets and open source code algorithms for the wildfire and flood risks, and (iv) the provision of an appropriate web-enabled multilingual secure user interface to enhance Civil Protection stakeholder experience.

\begin{figure}
\includegraphics[width=0.92\linewidth]{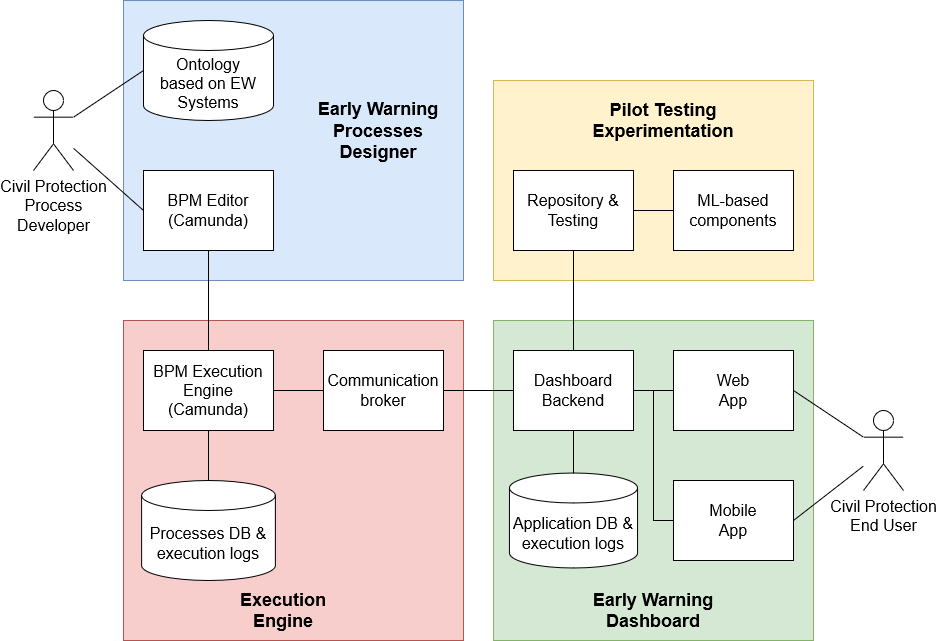}
\caption{TransCPEarlyWarning platform conceptual architecture.}
\label{fig1}
\end{figure}

The platform provides a unified solution that includes functionality for designing, executing and monitoring the stages of an early warning procedure. The system consists of three discrete subsystems that are interconnected through REST APIs. Those subsystems are in turn divided into smaller components that perform specific tasks. This modular design makes the system flexible and easy to upgrade, two qualities absolutely necessary when dealing with complex, dynamic, and diverse processes. As shown in Fig. \ref{fig1}, the three main subsystems, are the \textit{Early Warning Processes Designer}, the \textit{Execution Engine} and the \textit{Early Warning Dashboard}. 

The Early Warning Processes Designer is a tool that simplifies the design of an Early Warning Process and its main users will be Civil Protection officers. It provides  users with a simple and intuitive GUI that helps them design new processes or to transform the existing ones. An open source editor that can be used for the platform's purposes is Camunda's web editor \cite{camunda:online}. 

For providing guidance and insightful suggestions to the users, the subsystem utilises an ontology, named the Early Warning Ontology, which was created integrating the existing  different Civil Protection Early Warning processes and systems in the seven ADRION area countries. This Early Warning Ontology is a conceptual model that defines all the concepts concerning the activities and measures that are initiated from Civil Protection stakeholders. It defines the actors, that are different levels of national, regional or local authorities acting proactively for civil protection, the communications between them, the flow of the data, and the structure of the data such as the description of the exchanged messages. All these concepts are described with the formalism provided by OWL (Ontology Web Language) using the simplicity of defining classes, properties and relationships. This conceptual model is used to generate forms to help the process designers to create a valid Early Warning Business Process Model and Notation (BPMN) process for the proposed platform, tailored in the needs of each country. Additionally, it is used to define the integrated information that combines the domain knowledge of early warning systems.    

The users' input is then processed by the Execution Engine subsystem. The heart of this subsystem is the BPMN execution engine module. For this module, an existing open source system has been employed. Once again Camunda provides the required functionality and a REST API that enables its seamless embedding in the platform. Moreover, the BPMN execution engine subsystem also includes a Machine Learning module that automates tasks that require constant monitoring such as the early detection of a natural disaster. For the integration of the BPMN execution engine to the Machine Learning module and to the Early Warning Dashboard, a Communication broker was used. This module acts as an intermediate between all the other components and orchestrates the dissemination of the information. 

The third subsystem is the Early Warning Dashboard and can be viewed as the presentation layer that notifies the end users of the platform (state officers, volunteers etc.) and displays to them all the information they need to respond to a current event in a timely fashion. The Early Warning Dashboard consists of a back-end that is connected to the Execution Engine and feeds with the appropriate information the two front-end applications. These two applications provide the same functionality and differ only to the device they run on. The first one is a web application that is optimized for running on Desktop web browsers while the second one is a mobile application optimized for devices such as smartphone and tablets.

The Pilot Testing Experimentation component provides some additional capabilities to the platform as regards recent advancements in the early warning field using AI/ML. More specifically, it provides some basic functionality regarding the use of ML for detecting wildfires using visual data. It also provides a repository of existing algorithms and datasets in the field of early detection of wildfires, in order for stakeholders to familiarize themselves with such aspects, as well as to understand the current needs in the field, and how solutions could be applied in the Adriatic Ionian area.

\section{Integration with External Early Warning Systems}

At the early stages of the platform design, the definition of the Early Warning Ontology was used for modeling the complex and diverse early warning procedures that are employed by each country's Civil Protection. The goal of the system is to allow all involved parties to a Civil Protection process to reduce their response times and help them make better decisions by:
\begin{itemize}
    \item unifying all their sources of information,
    \item accelerating the dissemination of information,
    \item simplifying access to resources and documents describing the appropriate course of action.
\end{itemize}

Moreover, the conceptual model helped to create a context diagram that describes the role of all the involved parties and external bodies.

\begin{figure}
\includegraphics[width=0.92\linewidth]{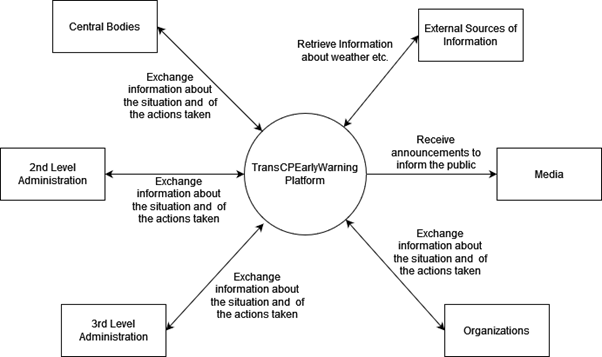}
\caption{TransCPEarlyWarning Context Diagram.}
\label{fig2}
\end{figure}

The involved parties and external bodies, as depicted in fig. \ref{fig2}, are the following:

\begin{itemize}
    \item \textbf{Central Bodies}: Central Bodies are responsible for monitoring and applying the Civil Protection processes at a country level.
    \item \textbf{2nd Level Administration}: The 2nd Level Administration is responsible for monitoring and applying the Civil Protection processes at a regional level.
    \item \textbf{3rd Level Administration}: The 3rd Level Administration is responsible for monitoring and applying the Civil Protection processes at a local level
    \item \textbf{Organizations}: With the general term `Organizations' we describe all those organizations (volunteers or not) that do not belong to one of the aforementioned categories.
    \item \textbf{Media}: Media describes the mass or social media used to inform the public.
    \item \textbf{External Sources of Information}: These are the systems used in every country for obtaining additional information that is useful in the decision-making process.
\end{itemize}

For interoperability purposes, the platform offers easy means of integration with external systems. The Early Warning Platform is designed to homogenize the alerts that are gathered by the Civil Protection mechanisms of each country and make them publicly available via an XML feed using the OASIS Standard Common Alerting Protocol - CAP (Version 1.2). CAP was selected because it is already used by organizations of the European Union, and it is perfectly suited for the needs of the platform, as it was specifically designed for publishing various types of alerts.

More specifically, the primary use of the CAP Alert Message is to provide a single input to activate all kinds of alerting and public warning systems.  This reduces the workload associated with using multiple warning systems while enhancing technical reliability and target-audience effectiveness. It also helps ensure consistency in the information transmitted over multiple delivery systems, another key to warning effectiveness.

Although primarily designed as an interoperability standard for use among warning systems and other emergency information systems, the CAP Alert Message can be used directly to alert recipients over various networks, including data broadcasts. Location-aware receiving devices can use the information in a CAP Alert Message to determine, based on their current location, whether that particular message was relevant to their users.

Moreover, the CAP Alert Message can also be used by sensor systems as a format for reporting significant events to collection and analysis systems and centers. This application is out of the scope of the designed platform but it is considered an important feature for future versions or systems integrating with it. Another potentially useful application of CAP is to normalize warnings from various sources so they can be aggregated and compared in tabular or graphic form as an aid to situational awareness and pattern detection.

\subsection{External CP Data Sources}

The proposed Civil Protection Early Warning platform needs to address the everyday operation of Civil Protection stakeholders in the ADRION area. In this context, it is a platform point of integration of existing systems/sources of information that are currently in use by the Civil Protection stakeholders in the area. Data about natural hazards, including floods and fires, are provided for Albania by the IGEO Institute of GeoSciences, as well as Meteo Tirana and MyDEWETRA. In Bosnia, the CP actors use the Network of European Meteorological Service, the Fire risk index and the the amount of precipitation from Federal Hydrometeorological Institute, and the water levels from Agency for watershed of Adriatic Sea. In Croatia, data is retrieved from Croatian Meteorological and Hydrological Service, and CROATIAN WATERS Display of water levels. The Hellenic national meteorological service, Lighting Maps, and Windy are some paradigms of the services used by the Greek CP authorities. Information systems of Administration of the republic of Slovenia for protection and rescue, Slovenian environmental agency (ARSO) and SMOK Monitoring, Warning and Control system are used in Slovenia. All these systems provide visualisation of data in tables and interactive maps. Some of them allow the data retrieval through REST APIs containing also GIS information. 

\section{Platform Security Specifications}

Due to the nature of the implemented platform, a strict approach was taken in the design and implementation of the security features of the platform that covers the whole life-cycle of the implemented services and software. The security specifications of the developed platform are focusing both at the system security and at the software architecture levels. Our approach for delivering a secure platform was based on the ``security by design'' principle, where security is considered at all phases of the software development process.

To facilitate the security requirements definition and analysis process, a security risk assessment was performed entailing the identification of critical assets for the platform, definition of security goals for each critical asset identified above and identification of threats to each security goal by identifying actors that may harm the platform and/or identifying sequences of actions that may result in intentional harm.

In terms of analysis and design, our focus was to ensure that the produced platform architecture meets all security, as well as the defined functional and non-functional requirements. At system security level, we employed practices that ``secure the perimeter'' of the platform to prevent malicious actors/input from crossing the boundary and entering the system from outside. This entails the combination of network and OS mechanisms, controls and protections, including firewalls and filters, intrusion detection/protection systems, as well as trending and monitoring of network traffic.

The security measures that were used for input and data validation and the control of internal processing work under the assumption that all input is malicious until proven otherwise, the need for validation of all input coming from sources (users, systems, databases, etc.) that are outside the platform's trust boundary, the assumption that client-side validation is not considered adequate and that server-side code will always perform its own validation. 

During platform implementation, a set of standard practices enhancing the security of the written code were followed. Such practices include minimising code size and complexity by using multiple small, simple, single-function modules instead of one large, complex module that performs multiple functions, implementing adequate fault handing (especially security-aware error and exception handling) and ensuring safe multitasking and multi threading. Moreover, to address significant security implications resulting from incorrect assumptions about the platform's execution environment, or from a misunderstanding of the interfaces between this system and its environment, our deployment approach was to ensure that the platform itself includes sufficient error and exception handling, even in the cases that it relies on specific execution environment components (middleware, OS, hardware platform, and network infrastructure) to perform security functions on its behalf, or to protect it from attack. This was necessary, in order to ensure that the platform is prevented from entering an insecure state, or having its security properties otherwise compromised, should any of the relied-on environment components fail. 

\section{Platform Implementation}
\subsection{User Interaction Dashboard}

The main goal of the platform's Dashboard is to provide a user-friendly way to monitor every step of the Early Warning Processes, retrieve all the necessary information to aid real-time decision making and orchestrate the execution of the tasks derived by those decisions. To achieve this objective, the Dashboard is compiled from a series of modules that implement the various tasks that are defined and automated by the BPMN execution engine. 

This modular approach is crucial to the viability of the platform, as it provides flexibility for implementing future Early Warning scenarios that may require functionality that was not foreseen. Moreover, the user experience is benefited by this approach because it allows the platform's administrators to select only the modules that are needed by each user and create personalized views that greatly simplify the user interface. 

Finally, the platform's Dashboard is assigned to control the type of information that is available to the users regarding their role and data access permissions. Users can be assigned to one or more groups that have been generated according to each country's CP plan and hierarchy of administration, thus, creating a complex yet elegant system for authorizing them to have access to multiple information levels. 

The basic modules that comprise the platform are briefly described below:

\begin{itemize}
\item \textbf{Task List}: This can be considered as the core module of the Dashboard and its goal is to inform users about the tasks they have to perform concerning an Early Warning procedure that has been initiated. It provides detailed information about these tasks and a button to notify the system that a task has been completed. If a task's actions can be partially or fully accomplished digitally, then appropriate web forms are displayed to the users helping them to execute these actions.
\item \textbf{Alerts}: The Alerts module is responsible for notifying the platform's users of an elevated fire or flood hazard in their respective region. Alerts are automatically disseminated through the BPMN execution engine but the users also have the option to forward them to other users or organizations, which are not participating in the predefined procedure. These external recipients can be stored and organized in user-defined lists for easier access.
\item \textbf{Danger Map}: This module presents a map, where each country's administrative subdivision is coloured in relation to the predicted danger for a natural disaster. This map offers a comprehensive way for the users to quickly determine if their region or neighbouring regions are at risk. Furthermore, it clarifies to all participants the exact borders of each region that sometimes may be hard to distinguish, especially in urban areas.
\item \textbf{Process Monitor}: The Process monitor informs the users of the status of an ongoing Early Warning procedure. Depending their position in the command hierarchy the users may see how far a process has been progressed and exactly which tasks have been completed.
\item \textbf{External Systems}: This module offers users the ability to combine multiple external monitoring systems in a single view. The users can create their virtual monitoring center by selecting publicly available systems they find useful from a wide list that is constantly updated. Subsequently, they can monitor them all at once in a unified view.
\end{itemize}

\subsection{Business Process Management Notation Execution Engine}

This part of the platform is responsible for the development and the execution of the business processes that describe the operational plans of the Civil Protection for each country. We chose the Business Process Model and Notation 2.0 (BPMN 2.0) \cite{bpmnspec:online} specification for modeling the workflows. 
BPMN 2.0 has become the de facto solution for modeling business processes since it is easily translatable to software components and at the same time it remains independent to specific development environments. Regarding the business execution engine, we have chosen Camunda \cite{camunda:online} which is an open-source workflow and decision-support platform. 
Camunda ships with several tools, like the Camunda Modeler which can be used for the modeling and the deployment of the processes to the business engine. Even though any third-party modeler, compliant with the BPMN 2.0 standard, can be used for the business process analysis, meaning the design of the operational aspects of the process, only Camunda Modeler can be utilized for the business process development, a task which involves the development of an executable process.

For fully exploiting the capabilities of Java, we have used Eclipse IDE \cite{eclipse22:online} for the process application development. The process application targets a shared Business Process Engine hosted in Apache Tomcat. It includes all necessary resources for the execution of the process, like BPMN files, HTML forms and Java delegates. We used ready project templates, which are called Archetypes, for instantiating the projects.

The execution of the business processes takes place in Camunda Business Process Engine, which runs as a standalone process and is accessible by a REST API. Its basic components are the following:
\begin{itemize}
\item Public API: It allows Java applications to interact with the process engine, features a command-style access pattern.
\item BPMN 2.0 Core Engine: it includes a lightweight execution engine for graph structures, a parser for transforming BPMN 2.0 files to java objects, and a set of BPMN 2.0 specific implementation.
\item Job Executor: this component deals with the asynchronous processing of background tasks like Timers.
\item Persistence: It includes a layer for storing process instance state to a relational database.
\end{itemize}

\subsection{AI-based Wildfire/smoke Detection System}

As regards the use cases for early warning systems based on AI, the most mature ones in terms of technologies and algorithms utilized, as well as number of deployments around the world, are the early warning systems used for wildfire and smoke detection. For this reason, we chose to focus on the use case of utilizing AI/ML-based techniques to detect wildfires and smoke, in order to produce the respective early warning events. Regarding the feasibility of integrating such a component to our platform, the following important aspects were considered:

\begin{enumerate}
\item Relevant AI/ML/DL algorithms availability: the first most fundamental requirement for implementing such an early warning system is the availability of algorithms specifically tackling this application domain.
\item Relevant dataset availability: the second most important aspect for such a system is to utilize visual data from actual wildfires and smoke in the environmental settings in which such events should be detected. Otherwise, the system will not be able to perform adequately.
\item Model training: having the algorithms and relevant data available allows for creating the respective AI model, i.e., the process to utilize data and train a model for the classification of visual data for wildfire and smoke.
\item Computing resources: cloud/server-based computing resources should be dedicated to execute the respective classification actions when visual data is available.
\end{enumerate}

Regarding the feasibility of integrating such algorithms and datasets for the system implemented in TransCPEarlyWarning EU project, there is a number of existing approaches available as open source, in the form of either code or datasets. There are also several representative papers with open-source implementations based on computer vision or AI techniques, the majority of which utilizes variations of well-known algorithms like the YOLO algorithm~\cite{yolov3}, and its newer versions. 

Regarding the availability of relevant datasets, apart from the ones mentioned above, the following datasets also can be utilized to support the development of an AI-based wildfire detection experimental prototype:

\begin{itemize}
\item The FLAME (Fire Luminosity Airborne-based Machine learning Evaluation) fire dataset~\cite{flame-dataset}.
\item The datasets of the University of Salerno for fire~\cite{salerno-fire} and smoke~\cite{salerno-smoke} detection.
\item The FIRESENSE database of videos for flame and smoke detection~\cite{firesense-db}.
\end{itemize}

Among these datasets, there is a variable degree of data annotations available to help develop a system for wildfire detection. Due to this, a certain amount of effort was channeled towards the annotation of such data sources, using tools such as \cite{labelimg} to manually annotate them. Regarding the availability of trained models, apart from code repositories and datasets, it appears to be limited currently, especially as regards models focusing on forest wildfires and smoke. For this reason, as regards the feasibility of developing a reliable system, we trained specific models for wildfire and smoke detection.

\section{Conclusion}
Early warning systems have been attracting a high degree of interest in the past few years, due to the growing rate of extreme weather events and other types of incidents linked to climate change, with their importance in the operation of civil protection systems growing fast. In this work, we presented a platform that provides the tools for supporting Civil Protection Early Warning in the Adriatic-Ionian area in Europe, stemming from the TransCPEarlyWarning EU project. Its main functionalities are related to the monitoring of Early Warning processes distributed among stakeholders from seven different countries in the area, at different levels of administration and integration of existing systems. The system also offers some basic functionality in terms of AI/ML-based detection case of wildfires. Regarding our future work, next steps will include the evaluation of the platform and the overall validation of the followed  design and implementation approach across the ADRION countries.

\bibliographystyle{IEEEtran}
\bibliography{TransCPEW}

\end{document}